\documentclass[twocolumn]{aastex701}

 \usepackage{amsmath}
\usepackage{booktabs}
\usepackage{array}
\usepackage{float}

\begin{document}


\title{Bimodality in Rotational Modulation of Planet-Hosting Stars}

\author[0000-0002-2106-4332]{Alexandre Araújo}
\affiliation{Centro de Rádio Astronomia e Astrofísica Mackenzie, Universidade Presbiteriana Mackenzie, Rua da Consolação, 930, SP, Brazil}
\email[show]{adesouza.astro@gmail.com}

\author[0000-0002-1671-8370]{Adriana Valio}
\affiliation{Centro de Rádio Astronomia e Astrofísica Mackenzie, Universidade Presbiteriana Mackenzie, Rua da Consolação, 930, SP, Brazil}
\email[]{adriana.valio@mackenzie.br}

\begin{abstract}

Stellar magnetic activity is governed by the interplay between rotation, convection, and the temporal evolution of surface magnetic structures, yet the role of planetary systems in shaping these processes remains unclear. In particular, it remains unclear whether the presence of planets can influence the global organization of stellar magnetic activity beyond localized or transient effects.
Here, we analyze Kepler photometry of more than 1,300 stars to investigate 
rotational modulation in stars with and without confirmed exoplanets. 
Using a time–frequency analysis, we measure a photometric proxy of rotational modulation dispersion, $S_{\rm phot}$, which traces the temporal coherence of surface magnetic features.
Stars with confirmed exoplanets exhibit systematically enhanced values of $S_{\rm phot}$ compared to stars without 
detected planets ($\Delta S_{\rm phot} = 0.17 \pm 0.01$ rad d$^{-1}$; 
$p < 10^{-25}$). 
More importantly, the $S_{\rm phot}$ distribution of planet hosts is bimodal, 
with peaks at $0.12$ and $0.44$ rad d$^{-1}$ (Hartigan's Dip Test 
$p < 10^{-6}$; $\Delta\mathrm{BIC}=188.7$), a feature absent in the 
control sample. We interpret $S_{\rm phot}$ as a proxy for rotational 
modulation dispersion, reflecting spot evolution rather than true 
differential rotation. The low-$S_{\rm phot}$ 
regime corresponds to stable magnetic coherence, or long-lived active regions ($\sim 6.8$ rotations), 
while the high-$S_{\rm phot}$ regime indicates rapidly evolving active 
regions ($\sim 0.9$ rotations).
The existence of two distinct regimes exclusively among planet-hosting stars suggests that planetary 
systems are associated with differences in the temporal organization of stellar magnetic activity. These results suggest that planets may influence stellar dynamos indirectly, by modifying the stability and evolution of surface magnetic structures rather than solely altering differential rotation.
\end{abstract}


\section{Introduction}

Stellar magnetic activity arises from the interaction between rotation, convection, and the evolution of surface magnetic structures \citep{Parker1955, Ossendrijver2003, Brun2017}. Photometric variability driven by starspots provides a powerful probe of these processes, encoding information about both stellar rotation and the temporal coherence of active regions \citep{Basri2011, mcquillan2014rotation}. 
Surface differential rotation, in particular, plays a central role in amplifying and reorganizing magnetic fields and is therefore a key parameter in models of stellar magnetism, activity cycles, and angular momentum evolution \citep{berdy+05}. 

However, the interpretation of this photometric variability is not straightforward: signals traditionally attributed to differential rotation can also be influenced by the finite lifetime and evolution of starspots, making it difficult to disentangle surface shear from magnetic-pattern evolution using light curves alone \citep{Aigrain2015,Basri2020}.
The frequency spread observed in time-frequency analyses of rotational modulation is typically dominated by spot evolution unless spot lifetimes exceed $\sim 10$ rotation periods 
\citep{basri2020information}.

Stars are usually assumed to evolve in isolation, with external agents exerting at most transient or localized influences on surface activity. Yet, a substantial fraction of stars host planetary systems, raising a fundamental question that  remains largely unexplored: can the presence of planets modify the global rotational dynamics of their host stars?
A growing body of theoretical work predicts that close-in planets can magnetically interact with their host stars through reconnection, Alfvénic coupling, or angular momentum exchange mediated by stellar winds \citep{Cuntz2000ApJ,Strugarek2016,
vidotto2025star}. Such star–planet interactions have been proposed as a source of localized chromospheric enhancements \citep{Shkolnik2003,lanza2008hot}, phase-dependent flaring, or excess coronal emission \citep{knutson2010correlation, araujo2021kepler, osborn2020investigating}. However, observational evidence for these effects has proven intermittent \citep{Pineda2026}  and often ambiguous, largely because stellar magnetic variability is intrinsically complex and time dependent. As a result, it remains unclear whether planets can exert a persistent influence on the global magnetic and dynamical properties of their host stars.

In this work, we define the observed frequency dispersion in stellar light curves as the range of dominant rotational frequencies.
We interpret this quantity as a photometric rotational shear proxy, denoted $S_{\rm phot}$, which captures the 
temporal structure of rotational modulation and serves primarily as a  diagnostic of the temporal coherence of the signal of surface magnetic activity. We investigate $S_{\rm phot}$ in a large sample of more than 1,300 stars observed by the Kepler mission \citep{borucki2010kepler}, comparing stars with confirmed planetary systems to those without detected planets.

In Section~\ref{sec:methods}, we describe the dataset, selection criteria, and the time–frequency methodology used to measure $S_{\rm phot}$ \citep{balona2016differential}. In Section~\ref{sec:results}, we present the main results of this work, beginning with the discovery of a bimodal $S_{\rm phot}$ distribution among planet-hosting stars and then exploring the population-level differences between stars with and without planets, as well as their dependence on stellar parameters. Finally, in Section~\ref{sec:conclusion}, we summarize our results and discuss their implications for stellar dynamo theory and star–planet interactions.

\section{Methods}\label{sec:methods}

We analyze a sample of 1,300 stars observed by the Kepler mission, divided into two subsets: stars with and without detected exoplanets. The  sample includes 809 stars hosting confirmed planets from the NASA Exoplanet Archive\footnote{https://exoplanetarchive.ipac.caltech.edu/} and 592 stars without detected planets from \cite{balona2016differential}. Both subsets span similar ranges in effective temperature and rotation period.

Our analysis is based on a time-frequency methodology \citep{balona2016differential} applied to Kepler light curves. The method tracks the temporal evolution of rotation signals from active regions by computing Lomb-Scargle periodograms in sliding windows and identifying the dominant frequency ridges. From these ridges, we measure the frequency spread $\Delta f = f_{\rm max} - f_{\rm min}$ and define the photometric rotational shear proxy:
\begin{equation}
S_{\rm phot} = 2\pi f_{\rm norm}\ \Delta f,
\end{equation}
where $f_{\rm norm} = 1.43$ is a normalization factor following \citet{balona2016differential}.
In addition to $S_{\rm phot}$, we compute complementary metrics, including the ridge frequency stability 
$\sigma_f$, that characterize the stability, persistence, and relative dispersion of the rotational modulation signal to ensure robust estimates (see Appendix \ref{ref:appendixA} and ~\ref{ref:appendixB}).

More importantly, as discussed in \cite{basri2020information}, photometric variability represents an integration of the stellar flux over the visible hemisphere and therefore does not uniquely map the underlying distribution of starspots. In this context, the observed frequency spread may arise from a combination of latitudinal differential rotation, spot evolution, finite spot lifetimes, and changes in active region configurations.

Throughout this work, we therefore interpret $S_{\rm phot}$ as a photometric rotational shear proxy, which captures the temporal dispersion of the rotational modulation. As emphasized by \citet{basri2020information, basri2022new}, unless spot lifetimes exceed $\sim 10$ rotation periods, the observed frequency spread is dominated by spot evolution rather than true latitudinal shear. Consequently, $S_{\rm phot}$ should be understood primarily as a diagnostic of the temporal coherence of surface magnetic activity, rather than a direct measurement of differential rotation.

\section{Results}\label{sec:results}

In this section, we present the main results of our analysis of the photometric rotational shear proxy, $S_{\rm phot}$, comparing stars with and without detected exoplanets. We first examine its global distribution and then explore population-level trends, correlations with stellar parameters, and the robustness of the observed features.
 
\subsection{A Bimodal Distribution of $S_{\rm phot}$ Exclusive to Planet Hosts}\label{sec:bimodal}

\begin{figure*}
\centering
\includegraphics[width=0.7\linewidth]{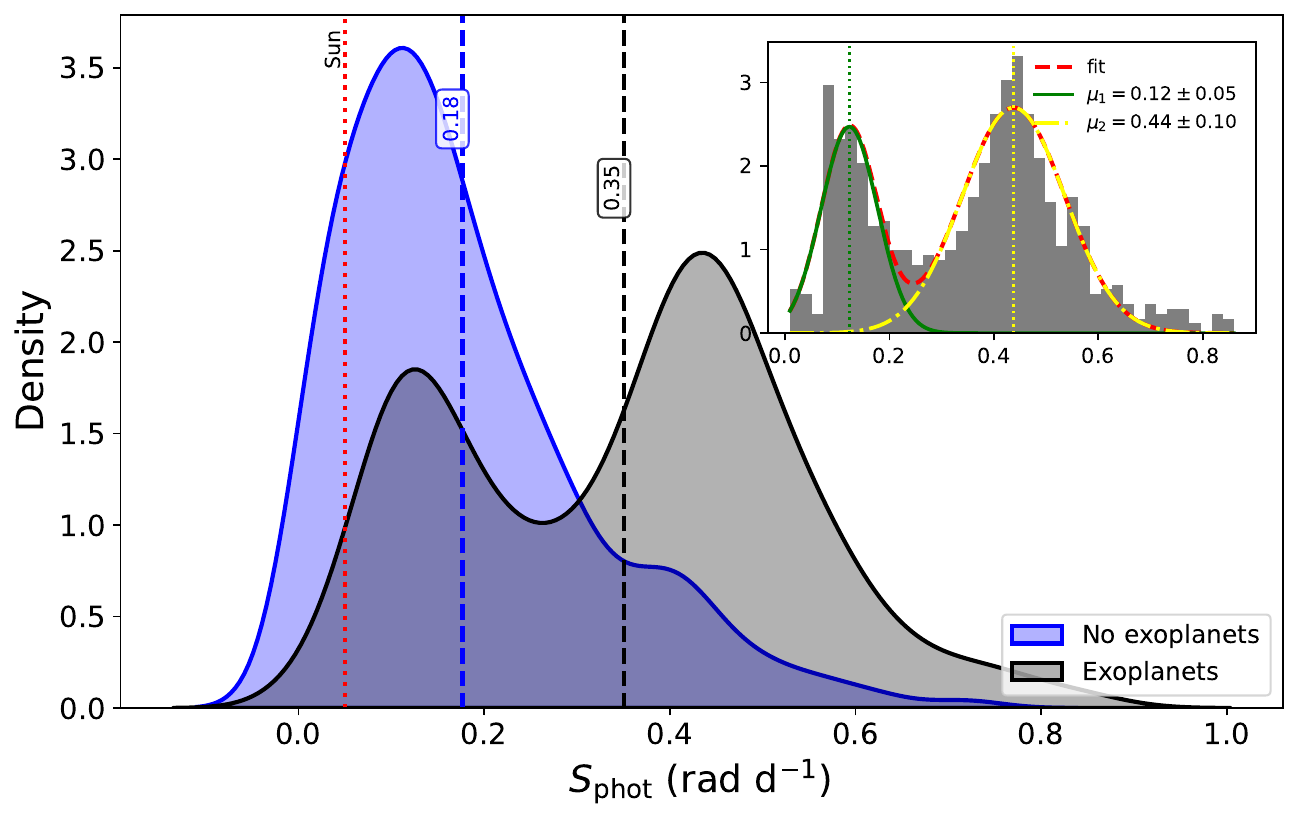}
\caption{Distribution of \textbf{$S_{\rm phot}$} for planet-hosting stars (gray) and stars without planets (blue). The inset shows the bimodal fit for planet hosts, with two well-separated Gaussian components.}
\label{fig:bimodal}
\end{figure*}

We begin by examining the distribution of the photometric rotational shear proxy, $S_{\rm phot}$, for the two stellar samples.
The  distribution for planet-hosting stars, shown in 
Figure \ref{fig:bimodal}, exhibits a clear bimodal structure with peaks at $0.12$ and $0.44$ rad\,d$^{-1}$. In contrast, this bimodality is not observed in the control sample.

To assess whether the observed distribution of $S_{\rm phot}$ is consistent with a single  population or requires multiple components, we applied complementary statistical tests of unimodality and model selection. Hartigan's Dip Test \citep{Hartigan1985}, which quantifies deviations from a unimodal distribution without assuming any specific functional form, strongly rejects the null hypothesis of unimodality ($p < 10^{-6}$). 

To further characterize the structure of the distribution, we performed Gaussian Mixture Modeling (GMM), fitting models with different numbers of components and comparing them using the Bayesian Information Criterion (BIC), which penalizes model complexity. The two-component model is strongly favored over a single-component description ($\Delta\mathrm{BIC}_{1\rightarrow2}=188.7$), indicating that the data are better represented by two distinct subpopulations. 

Furthermore, we tested the complementary metrics defined in Appendix ~\ref{ref:appendixA} for bimodality. The ridge frequency stability, $\sigma_f$, exhibits a bimodal distribution exclusively among planet hosts (Dip Test $p = 0.006$; $\Delta\mathrm{BIC} = 349.3$), while the control sample remains consistent with a unimodal distribution ($p = 0.945$). Similarly, the relative dispersion $S_{\rm phot}/\Omega$ shows significant bimodality in planet hosts ($p = 0.001$; $\Delta\mathrm{BIC} = 205.9$) but not in the control sample ($p = 0.743$). The consistency of bimodality across three independent metrics---$S_{\rm phot}$, $\sigma_f$, and relative dispersion---demonstrates that the signal is intrinsic to the temporal organization of surface magnetism and not an artifact of any single measurement definition.

To ensure that this result is not driven by underlying correlations with stellar parameters, we repeated the analysis after removing the dependence on rotation period; the preference for bimodality remains significant ($\Delta\mathrm{BIC}_{\rm residuals}=109.4$). Finally, we evaluated the robustness of the bimodal structure through Monte Carlo resampling, perturbing the data within their measurement uncertainties. In all realizations, the two-component model is preferred ($\Delta\mathrm{BIC} > 10$), with a mean $\Delta\mathrm{BIC} = 131.9$, demonstrating that the bimodality is not an artifact of observational noise but a stable feature of the data.

\begin{table*}
\footnotesize
\centering
\caption{Stellar and planetary statistical properties associated with the two peaks of the bimodal \textbf{$S_{\rm phot}$} distribution.}\label{tab:stellar_planetary_deltaomega}
\begin{tabular}{lccccccccc}
\toprule
\multicolumn{9}{c}{\textbf{Stellar Properties}} \\
\hline
\textbf{$S_{\rm phot}$} peak & $N_\star$ & $\langle S_{\rm phot}\rangle$ &
$\sigma_{S_{\rm phot}}$ & $\langle P_{\rm rot}\rangle$ &
$\sigma_{P_{\rm rot}}$ & $\langle T_{\rm eff}\rangle$ &
$\sigma_{T_{\rm eff}}$ & $\langle R_\star\rangle$ & \textbf{$\langle\sigma_f\rangle$} \\
 &  & (rad d$^{-1}$) & (rad d$^{-1}$) & (d) & (d) & (K) & (K) & ($R_\odot$) & \textbf{(c\,d$^{-1}$)} \\
\hline
High-\textbf{$S_{\rm phot}$} & 565 & 0.44 & 0.111 & 20.17 & 9.53 & 5069 & 630 & 0.845 & \textbf{0.0123} \\
Low-\textbf{$S_{\rm phot}$}  & 244 & 0.12 & 0.062 & 13.94 & 5.86 & 5172 & 632 & 0.870 & \textbf{0.0024} \\
\hline
\multicolumn{10}{c}{\textbf{Planetary Properties}} \\
\hline
\textbf{$S_{\rm phot}$} peak & $N_p$ & $\langle P_{\mathrm{orb}}\rangle$ &
$\sigma_{P_{\mathrm{orb}}}$ & $\langle R_p\rangle$ &
$\sigma_{R_p}$ & $\langle [\mathrm{Fe/H}]\rangle_\mathrm{host}$ & \multicolumn{2}{c}{Fraction [Fe/H] $< -0.2$} & \textbf{Coherence} \\
 &  & (d) & (d) & ($R_{\mathrm{Jup}}$) & ($R_{\mathrm{Jup}}$) & & (\%) & & \textbf{(rotations)} \\
\hline
High-\textbf{$S_{\rm phot}$} & 808 & 26.00 & 58.92 & 0.222 & 0.157 & $0.0028$ & \multicolumn{2}{c}{$9.8\%$} & \textbf{0.9} \\
Low-\textbf{$S_{\rm phot}$}  & 332 & 25.67 & 65.02 & 0.248 & 0.212 & $0.0277$ & \multicolumn{2}{c}{$2.5\%$} & \textbf{6.8} \\
\hline
\multicolumn{10}{c}{\textbf{Correlations with \textbf{$S_{\rm phot}$} (Spearman's $\rho$)}} \\
\hline
Peak & \multicolumn{2}{c}{\textbf{$S_{\rm phot}$} $\times P_{\mathrm{orb}}$} &
\multicolumn{2}{c}{\textbf{$S_{\rm phot}$} $\times R_p$} &
\multicolumn{2}{c}{\textbf{$S_{\rm phot}$} $\times [\mathrm{Fe/H}]$} &
\multicolumn{2}{c}{\textbf{$S_{\rm phot}$} $\times \mathrm{Age}$} &
\multicolumn{1}{c}{\textbf{$S_{\rm phot}$} $\times \sigma_f$} \\
\hline
High-\textbf{$S_{\rm phot}$} & \multicolumn{2}{c}{$0.067$ ($p=0.16$)} &
\multicolumn{2}{c}{$0.062$ ($p=0.15$)} &
\multicolumn{2}{c}{$-0.041$ ($p=0.45$)} &
\multicolumn{2}{c}{$0.125$ ($p=0.10$)} &
\multicolumn{1}{c}{\textbf{$0.31$ ($p = 10^{-12}$)}} \\
Low-\textbf{$S_{\rm phot}$}  & \multicolumn{2}{c}{$-0.039$ ($p=0.58$)} &
\multicolumn{2}{c}{$0.063$ ($p=0.38$)} &
\multicolumn{2}{c}{$0.020$ ($p=0.79$)} &
\multicolumn{2}{c}{$-0.037$ ($p=0.72$)} &
\multicolumn{1}{c}{\textbf{$0.81 (p = 10^{-88})$}} \\
Planet Hosts (All) & \multicolumn{2}{c}{$0.041$ ($p=0.30$)} &
\multicolumn{2}{c}{$0.015$ ($p=0.70$)} &
\multicolumn{2}{c}{$-0.029$ ($p=0.51$)} &
\multicolumn{2}{c}{$0.127$ ($p=0.037$)} &
\multicolumn{1}{c}{\textbf{$0.79$ ($p=10^{-314}$)}} \\
No planet (All) & \multicolumn{2}{c}{--} &
\multicolumn{2}{c}{--} &
\multicolumn{2}{c}{--} &
\multicolumn{2}{c}{--} &
\multicolumn{1}{c}{\textbf{$0.70$ ($p=10^{-88}$)}} \\
\botrule
\end{tabular}

\vspace{0.2cm}
\begin{minipage}{\linewidth}
\footnotesize
\textit{Notes.} $S_{\rm phot}$ (in rad\,d$^{-1}$).
$N_\star$ and $N_p$ represent the number of stars and planets associated with each 
\textbf{$S_{\rm phot}$} peak, respectively. For planetary properties, $N_p = 1140$ total planets (808 in high peak, 332 in low peak), with radius measurements available for $N=655$ planets. 
The table lists the mean values and corresponding standard deviations for 
\textbf{$S_{\rm phot}$}, stellar rotation period ($P_{\rm rot}$, in days), effective temperature 
($T_{\rm eff}$, in Kelvin), stellar radius ($R_\star$, in solar radii), ridge frequency stability ($\sigma_f$, in c\,d$^{-1}$), planetary orbital period 
($P_{\mathrm{orb}}$, in days), planetary radius ($R_p$, in Jupiter radii), magnetic coherence timescale (in rotation periods), and host star metallicity ([Fe/H]). 
The fraction of metal-poor stars ([Fe/H] $< -0.2$) is based on $N=501$ stars with metallicity measurements (338 high, 163 low). 
``High'' and ``Low'' \textbf{$S_{\rm phot}$} peaks correspond to the high- and low-shear components 
of the bimodal \textbf{$S_{\rm phot}$} distribution. All correlations are Spearman's rank correlation coefficients with associated $p$-values. The strong correlation between $S_{\rm phot}$ and $\sigma_f$ ($\rho = 0.79$) confirms that $S_{\rm phot}$ is fundamentally a measure of rotational modulation dispersion.
\end{minipage}
\end{table*}

Observational effects may influence the measured amplitudes of $S_{\rm phot}$. In particular, transiting planets are preferentially detected in nearly edge-on orbits, and if the stellar spin axis is approximately aligned with the orbital plane, the stellar rotation axis is oriented close to perpendicular to the line of sight. In this configuration, starspots at different latitudes produce more distinguishable rotational signals in photometric data, improving the sensitivity to rotational modulation dispersion. Conversely, for stars observed at low inclination, latitudinal differences are foreshortened, leading to a systematic underestimation of $S_{\rm phot}$. This geometric bias can therefore enhance the detectability of rotational shear in planet-hosting stars and shift the measured distribution toward higher values. However, such an effect is expected to produce a continuous bias rather than a discrete bimodal structure, and thus cannot by itself account for the presence of two distinct coherence regimes observed exclusively among planet-hosting stars.

Moreover, the detectability of transiting exoplanets is favored around stars with lower photometric variability and reduced magnetic activity \citep{Ciardi2011, Gilliland2011}, which could, in principle, bias samples toward systems with weaker surface shear. However, the observed enhancement of $S_{\rm phot}$ in planet-hosting stars runs counter to this expectation, since higher $S_{\rm phot}$ may enhance toroidal field generation and thus potentially increase magnetic activity \citep{Brun2017}. In particular, such biases offer no natural explanation for the emergence of a bimodal $S_{\rm phot}$ distribution that is observed exclusively among stars with planets (Figure \ref{fig:bimodal}).

\subsection{Population-Level Differences}
Having established that planet-hosting stars exhibit a bimodal distribution in $S_{\rm phot}$, we now investigate how these stars differ from the comparison sample in terms of global stellar properties and correlations with stellar parameters.

\begin{figure*}
\centering
\includegraphics[width=1\textwidth]{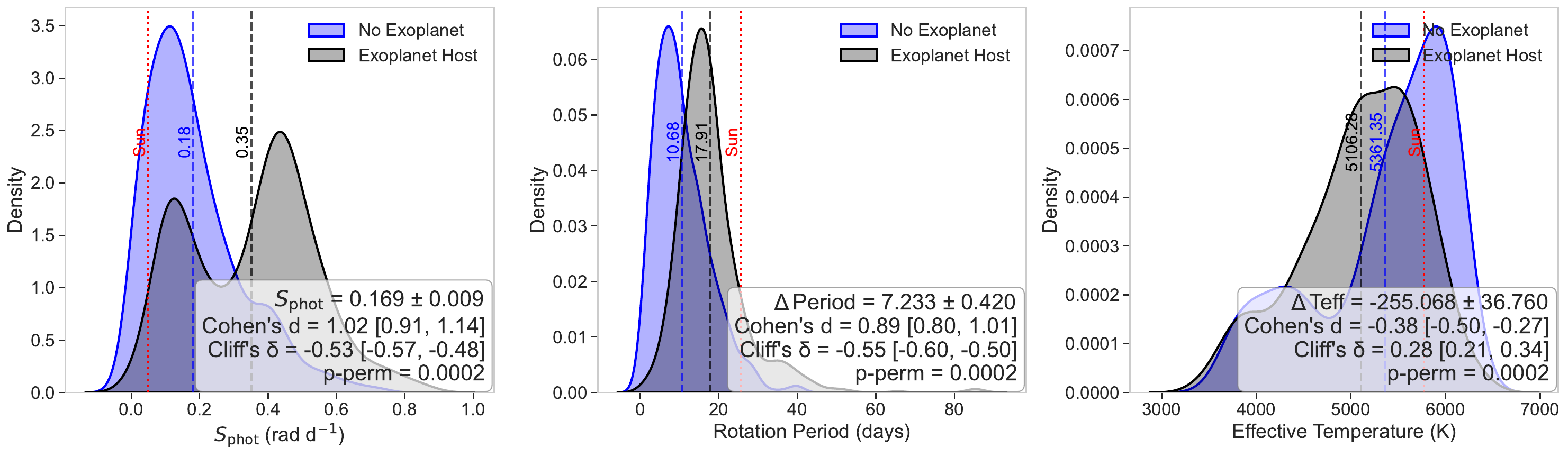}
\caption{
Distributions of  $S_{\rm phot}$ (left), rotation period (center), and effective temperature (right) for stars with (gray) and without (blue) detected exoplanets.
Planet-hosting stars exhibit enhanced $S_{\rm phot}$, slower rotation, and lower temperatures on average.
}
\label{fig:densidade}
\end{figure*}

The distributions of \textbf{$S_{\rm phot}$}, rotation period, and effective temperature for stars with and without detected exoplanets are shown in Figure~\ref{fig:densidade}. The left panel reveals that planet-hosting stars (gray) exhibit a systematic shift toward higher \textbf{$S_{\rm phot}$} values, with a mean difference of $0.17 \pm 0.01$ rad\,d$^{-1}$ relative to stars without planets (purple).

Moreover, as shown in the middle panel of Figure~\ref{fig:densidade}, planet-hosting stars also rotate more slowly (longer rotational periods) and are, on average, cooler than stars without detected planets. This indicates that population effects may influence the raw distributions and must be carefully controlled.

A limitation of our control sample is that stars classified as ``without detected planets'' may in fact host planetary systems that remain undetected by transit photometry, either due to unfavorable orbital inclinations or sensitivity thresholds of the Kepler mission. If the hypothesis presented in this work is correct — namely, that the presence of planets is associated with enhanced \textbf{$S_{\rm phot}$} — this contamination artificially elevates the mean \textbf{$S_{\rm phot}$} of the control group, rendering the detection of the effect more conservative. The fact that we observe statistically significant differences despite this bias suggests that the underlying physical effect may be even more pronounced than reported here.

\subsection{Multivariate Control for Stellar Parameters}

To test whether the enhancement in \textbf{$S_{\rm phot}$} can be attributed solely to differences in $T_{\mathrm{eff}}$ or $P_{\mathrm{rot}}$, we performed a multiple linear regression:
\begin{equation}
S_{\rm phot} = \beta_0 + \beta_1 T_{\mathrm{eff}} + \beta_2 P_{\mathrm{rot}} + \beta_3 I_{\mathrm{planet}},
\end{equation}

\noindent where $I_{\mathrm{planet}}$ is a binary indicator identifying stars hosting confirmed exoplanets.
A model including only $T_{\mathrm{eff}}$ and $P_{\mathrm{rot}}$ explains 30\% of the variance ($R^2 = 0.30$). 
Adding the planetary indicator  improves the fit ($R^2 = 0.40$; $\Delta$BIC $=119$), with 
$\beta_3 = 0.109 \pm 0.009$ rad\,d$^{-1}$ ($p < 10^{-25}$). 

This result indicates that the presence of planets is statistically associated with enhanced \textbf{$S_{\rm phot}$}, even after accounting for the population-level differences seen in Figure~\ref{fig:densidade}. 
The magnitude of this coefficient represents a substantial fraction of the \textbf{$S_{\rm phot}$} amplitudes observed in the sample, indicating that the planetary contribution is not negligible compared to intrinsic stellar variations.

\begin{figure*}
    \centering
    \includegraphics[width=0.7\linewidth]{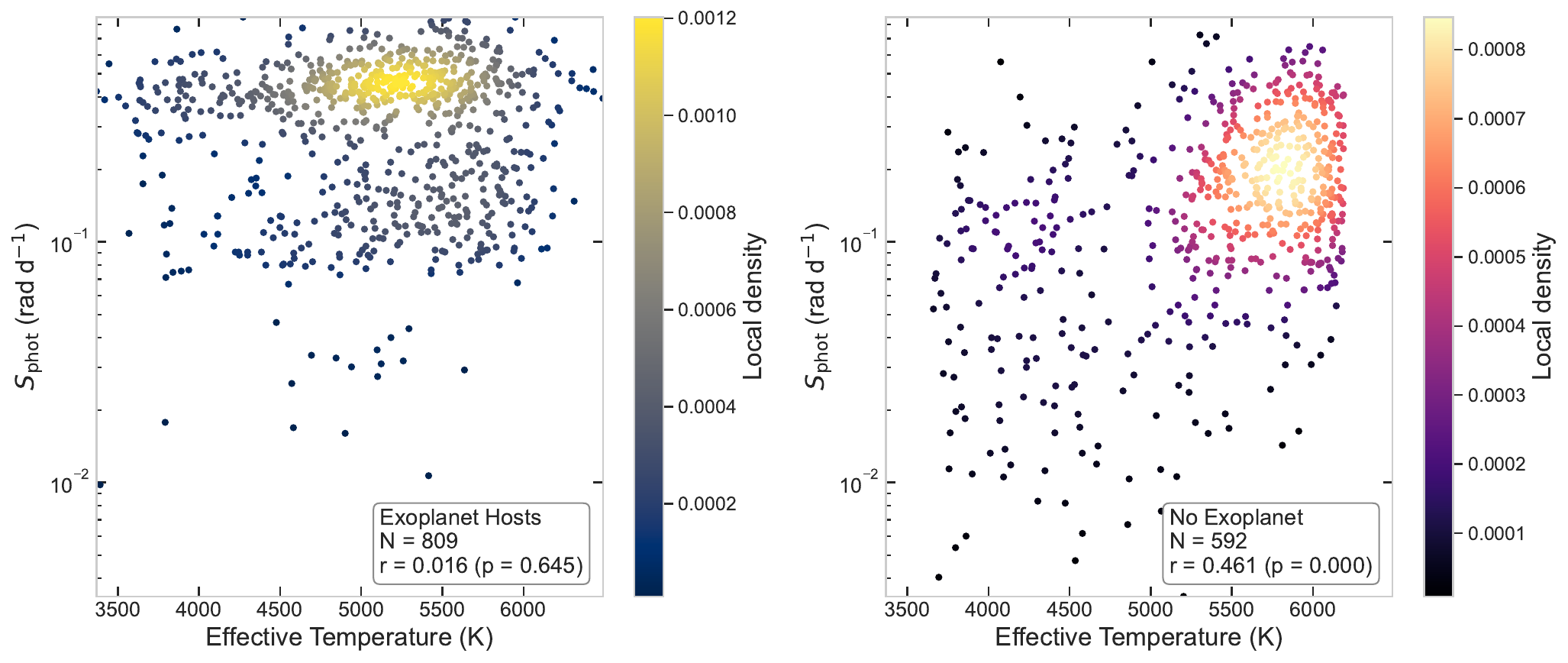}
    \includegraphics[width=0.7\linewidth]{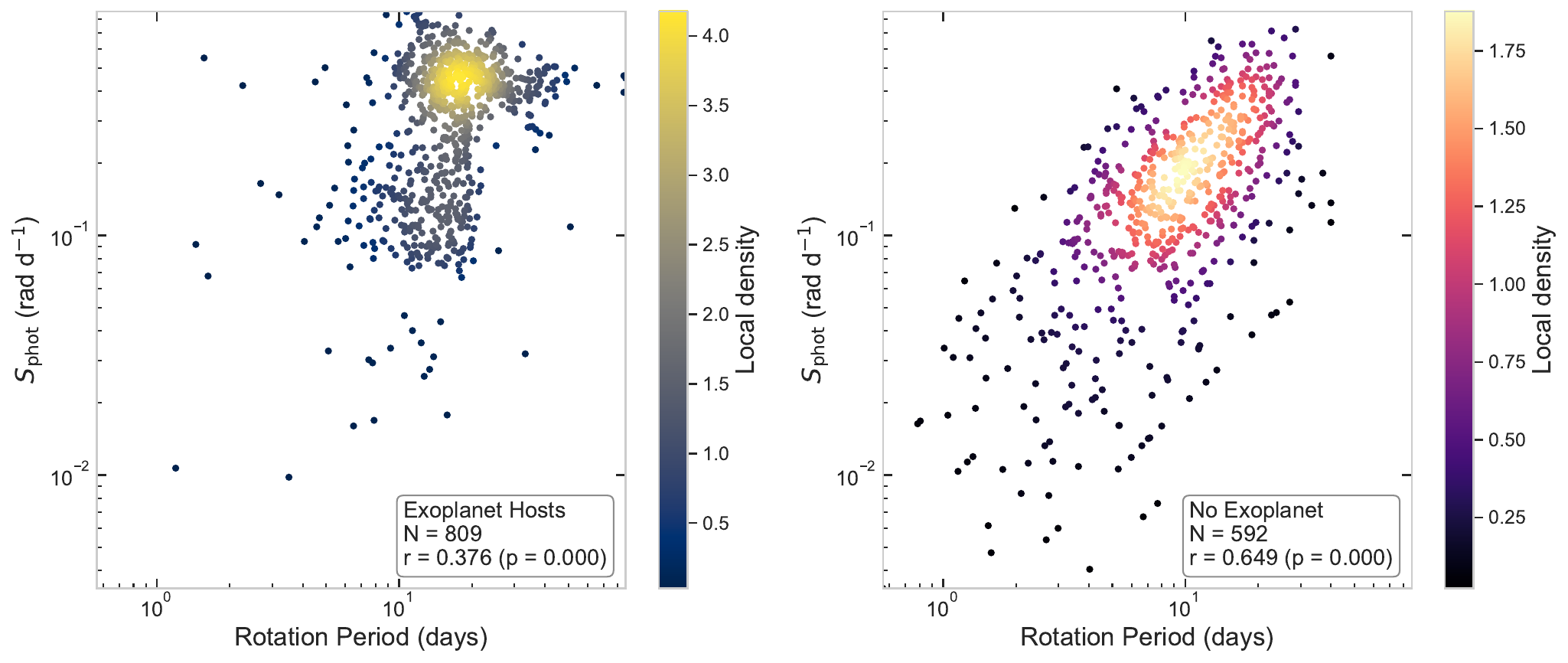}
    \caption{{$S_{\rm phot}$} versus effective temperature (top) and rotation period (bottom) for exoplanet hosts (left) and stars with non-detected planets (right). Color intensity represents local point density. Correlation coefficients ($r$) and $p$-values are indicated.}
    \label{fig:correlation}
\end{figure*}

Figure~\ref{fig:correlation} shows the correlation of \textbf{$S_{\rm phot}$} with the effective temperature ($T_{\rm eff}$) and rotation period ($P_{\rm rot}$) for two distinct samples: planet-hosting stars (left panels) and stars without detected planets (right panels).
The top panels of Figure~\ref{fig:correlation} present \textbf{$S_{\rm phot}$} as a function of $T_{\rm eff}$ and reveal a clear difference between the two samples. For planet-hosting stars, we find no significant correlation, with $r = 0.016$ ($p = 0.645$), indicating that \textbf{$S_{\rm phot}$} is essentially independent of effective temperature for this population. In contrast, stars without detected planets exhibit a moderate positive correlation, with $r = 0.461$ ($p < 0.001$).

This disparity suggests fundamental differences between the two populations. The lack of correlation for planet hosts may indicate that the presence of planetary systems alters the relationship between rotational modulation dispersion and stellar temperature, possibly through star--planet interactions or through selection effects related to planet detectability. The moderate correlation for non-host stars is consistent with theoretical expectations linking convective properties to effective temperature \citep{Kitchatinov2011}.

The bottom panels of Figure~\ref{fig:correlation} show \textbf{$S_{\rm phot}$} as a function of the rotation period. Both samples exhibit positive correlations, but with markedly different strengths. For planet hosts, we find $r = 0.376$ ($p < 0.001$), while non-host stars show a stronger correlation of $r = 0.649$ ($p < 0.001$).

The positive correlations indicate that stars with longer rotation periods (slower rotators) tend to exhibit larger \textbf{$S_{\rm phot}$} values. This trend is consistent with the interpretation of $S_{\rm phot}$ as a diagnostic of rotational modulation dispersion: slower rotators may show more complex spot patterns or longer spot lifetimes, leading to broader frequency spreads in the time-frequency domain.

However, as emphasized by \citet{basri2020information}, theoretical models predict a complex dependence on rotation through the Rossby number, and the observed correlation should not be interpreted as a simple monotonic relation with true differential rotation \citep{kuker2011differential, Kitchatinov2011, Brun2017}.

The weaker correlation for planet hosts suggests that the presence of planets may introduce additional scatter or modify the rotational evolution of their host stars. The higher correlation coefficient for non-host stars likely represents the intrinsic relationship between rotation period and rotational modulation dispersion in the absence of planetary companions.

\subsection{Star--Planet Interaction and Bimodality}

Considering that $S_{\rm phot}$ is interpreted primarily as a diagnostic of magnetic activity coherence rather than kinematic differential rotation, our results reveal two distinct regimes in the temporal organization of surface magnetic activity among planet-hosting stars. These regimes may arise from differences in stellar dynamo operation or magnetic field organization.

The low-$S_{\rm phot}$ regime ($S_{\rm phot} \lesssim 0.2$ rad\,d$^{-1}$) is characterized by stable rotational modulation, with low ridge frequency dispersion ($\sigma_f \sim 0.002$ cycle d$^{-1}$) and longer magnetic coherence timescales ($\sim 6.8$ rotation periods). This regime is consistent with active regions that maintain their phase coherence over multiple rotations. In contrast, the high-$S_{\rm phot}$ regime ($S_{\rm phot} \gtrsim 0.3$ rad\,d$^{-1}$) is associated with unstable rotational modulation, exhibiting higher $\sigma_f$ ($\sim 0.012$ cycle d$^{-1}$) and very short coherence timescales ($\sim 0.9$ rotation periods), corresponding to rapidly evolving active regions where the rotational signal loses coherence quickly.

This interpretation may also connect with the phenomenology described by \cite{basri2022new}, who analyzed spot lifetimes through the persistence of cross-correlation peaks across different temporal windows. Although our methodology is fundamentally different and does not apply the \cite{basri2022new} framework directly, both approaches probe the temporal coherence of rotational modulation. In our case, the sliding Lomb–Scargle analysis measures the dispersion and stability of the dominant rotational frequency ridge, which may similarly be affected by spot evolution and finite spot lifetimes. We performed a preliminary cross-match using the quality criteria required for a Basri-type spot lifetime analysis and found that only $\sim$20 stars in our sample satisfy those constraints, preventing a statistically meaningful comparison in the present work. A dedicated investigation connecting $S_{\rm phot}$, spot lifetimes, and cross-correlation persistence metrics will therefore be pursued in future studies.

To further explore this connection, we examine the stellar and planetary properties associated with each $S_{\rm phot}$ regime, summarized in Table~\ref{tab:stellar_planetary_deltaomega}. Stars in the low-\textbf{$S_{\rm phot}$} regime (\textbf{$S_{\rm phot}$} $\lesssim 0.2$ rad\,d$^{-1}$) rotate  faster on average than those in the high-\textbf{$S_{\rm phot}$} regime (\textbf{$S_{\rm phot}$} $\gtrsim 0.3$ rad\,d$^{-1}$) ($\langle P_{\rm rot}\rangle = 13.9$ d versus $20.2$ d; Mann-Whitney $p < 10^{-4}$), while their effective temperatures are statistically indistinguishable.

Moreover, no statistically significant correlations are found between \textbf{$S_{\rm phot}$} and planetary properties such as orbital period or planetary radius, either for the full sample or within each peak (Table~\ref{tab:stellar_planetary_deltaomega}). This suggests that the bimodality is a global property of the host star, rather than a direct function of the planetary characteristics. While magnetic star–planet interactions could in principle influence stellar activity, the absence of correlations with individual planetary parameters indicates that any such effect must operate through global or threshold-dependent processes, possibly linked to magnetic coupling between the star and the planetary system, rather than through localized or continuously varying interactions.

\subsection{Stability of the rotational modulation, $\sigma_f$}
Table~\ref{tab:stellar_planetary_deltaomega} reveals a strong positive correlation between $S_{\rm phot}$ and $\sigma_f$ ($\rho = 0.79$, $p = 10^{-314}$). Since $\sigma_f$ measures the stability of the rotational frequency ridge---a direct observable independent of any conversion to shear units---this correlation provides strong evidence that $S_{\rm phot}$ is primarily a measure of rotational modulation dispersion. The tight relationship between these two metrics confirms that the observed $S_{\rm phot}$ enhancement and bimodality reflect genuine differences in the temporal coherence of surface magnetic activity.

The robustness of the bimodality is further reinforced by the independent metric $\sigma_f$, which measures the stability of the rotational frequency ridge without conversion to shear units. The $\sigma_f$ distribution is bimodal exclusively in planet hosts (Dip Test $p = 0.006$; $\Delta\mathrm{BIC} = 349.3$), while the control sample remains unimodal ($p = 0.945$). The consistency of bimodality across $S_{\rm phot}$, $\sigma_f$, and relative dispersion demonstrates that the signal is intrinsic to the temporal coherence of rotational modulation and not an artifact of any single measurement definition.

\section{Discussion and Conclusion}\label{sec:conclusion} 
We show that stars hosting exoplanets exhibit systematically enhanced values of the photometric rotational modulation dispersion, $S_{\rm phot}$, compared to stars without detected planets. This difference remains significant even after controlling for stellar effective temperature and rotation period (Appendix~\ref{ref:appendixB}).

A key result of this work is that the $S_{\rm phot}$ distribution of planet-hosting stars is bimodal, with two well-separated regimes that are absent in the comparison sample. 
An intriguing possibility is that the two peaks of the bimodal distribution correspond to stars in fundamentally different magnetic activity states. Stars in the low-$S_{\rm phot}$ regime (stable modulation, high coherence) may operate in a dynamo configuration where active regions are long-lived and maintain phase coherence over many rotations. Conversely, stars in the high-$S_{\rm phot}$ regime (unstable modulation, low coherence) may represent systems where rapid spot evolution and complex magnetic topologies produce irregular rotational signals. This hypothesis is currently being investigated through detailed magnetic activity indicators and will be the subject of future studies.

The absence of significant correlations between $S_{\rm phot}$ and planetary properties suggests that this bimodality is a global property of the host stars rather than a direct response to individual planetary characteristics. While the underlying physical mechanism remains uncertain, the results are consistent with a possible scenario in which planetary systems are associated with long-term modifications of the temporal organization of surface magnetism. The present interpretation should nevertheless be regarded as exploratory. Thus, the bimodality may reflect stars operating in distinct regimes of magnetic activity coherence, potentially influenced by star--planet interactions.

In this framework, star--planet interactions may affect the large-scale magnetic environment by modifying the lifetime, stability, or longitudinal organization of active regions, thereby shifting stars between distinct regimes of photometric variability. 
This interpretation is consistent with a scenario in which planetary systems affect the magnetic dynamo indirectly, through changes in the temporal structure of surface magnetism rather than through a direct modification of latitudinal shear \citep{vidotto2025star}.

In summary, these findings suggest that planetary systems may play a role in shaping the temporal coherence of stellar magnetic activity and, by extension, stellar magnetic dynamos. The enhanced $S_{\rm phot}$ and distinct bimodality in planet-hosting stars may reflect altered modes of dynamo operation, with direct consequences for stellar activity cycles, angular momentum loss, and the long-term magnetic environment experienced by orbiting planets. 

\begin{acknowledgments}
We thank L. A. Balona for kindly sharing the data used in this work. We also thank the anonymous referee for a careful reading of the manuscript and for constructive comments that helped improve the interpretation of the results.
The authors acknowledge partial funding from Brazilian agencies FAPESP (grant \#2021/02120-0) and CNPq (grants \#150817/2022-3, and \#172886/2023-6). 
\end{acknowledgments}

\appendix
\twocolumngrid

\section{Quality criteria for S$_{phot}$ estimates and additional metrics.} 

\begin{table}[h]
\centering
\caption{Quality criteria for $S_{\rm phot}$.}
\label{tab:quality}
\begin{tabular}{l c l}
\hline
Criterion & Threshold & Description \\
\hline
$N_{\rm windows}$ & $\geq 20$ & Temporal persistence \\
$\sigma_f$ & $\leq 0.05$ c\,d$^{-1}$ & Ridge coherence \\
SNR & $\geq 5$ & Peak significance \\
RMS & $\geq 5\times10^{-4}$ & Variability level \\
\hline
\end{tabular}
\end{table}

\label{ref:appendixA}
In addition to $S_{\rm phot}$, we compute complementary metrics to characterize the rotational modulation:
\begin{itemize}
\item $\sigma_f$: standard deviation of the dominant ridge frequency, measuring the stability of  rotational modulation;
\item $N_{\rm windows}$: number of time windows in which the rotational signal is detected, quantifying temporal persistence;
\item Relative dispersion: $S_{\rm phot}$, where $\Omega = 2\pi/P_{\rm rot}$, providing a dimensionless measure of rotational modulation spread.
\end{itemize}
These metrics, particularly $\sigma_f$, provide independent diagnostics of magnetic activity coherence that do not rely on the conversion to $S_{\rm phot}$.

\section{Robustness to Stellar Parameters} \label{ref:appendixB}

\begin{figure*}
\centering
\includegraphics[width=0.85\textwidth]{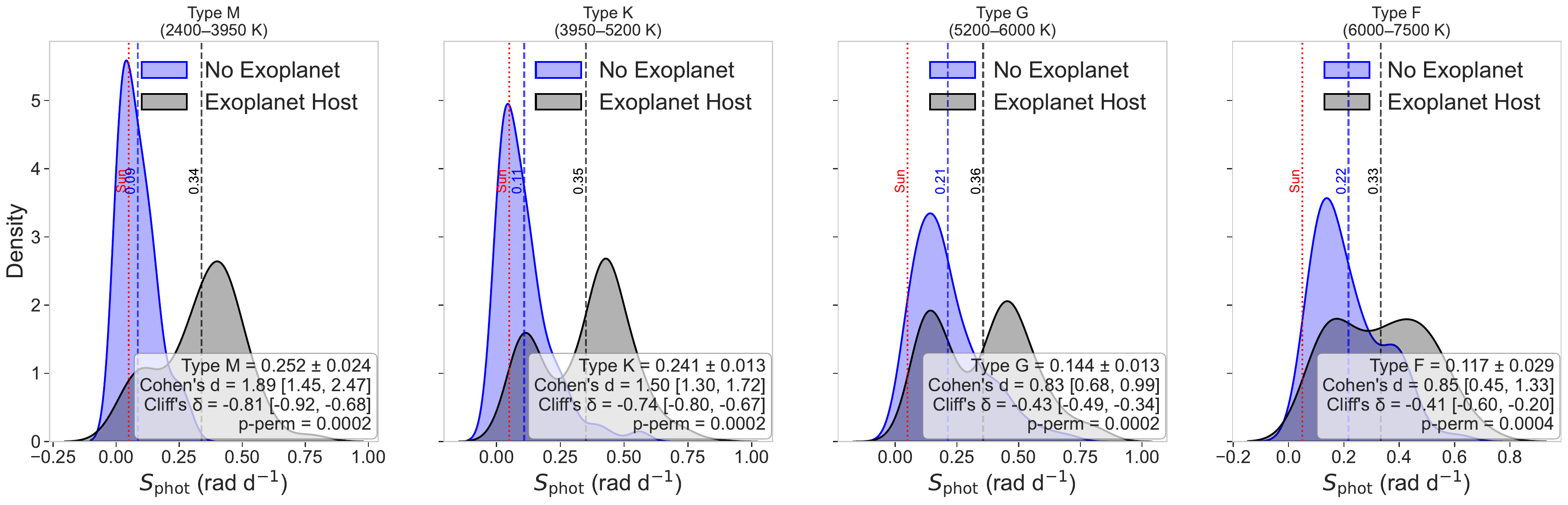}
\includegraphics[width=0.85\textwidth]{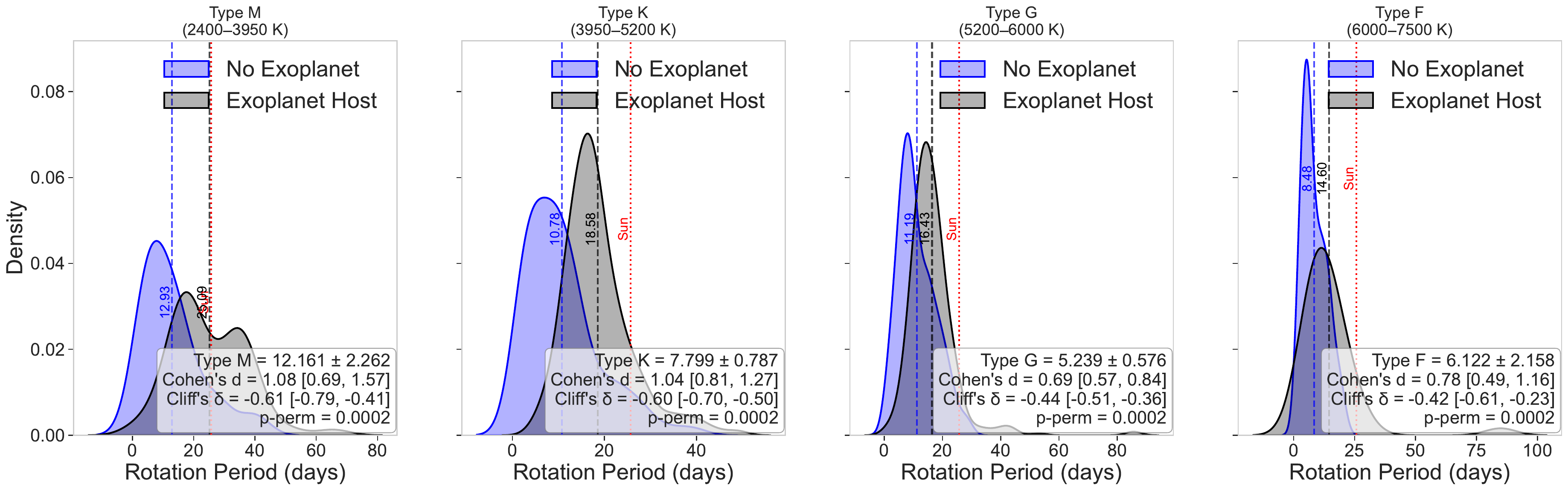}
\includegraphics[width=0.85\textwidth]{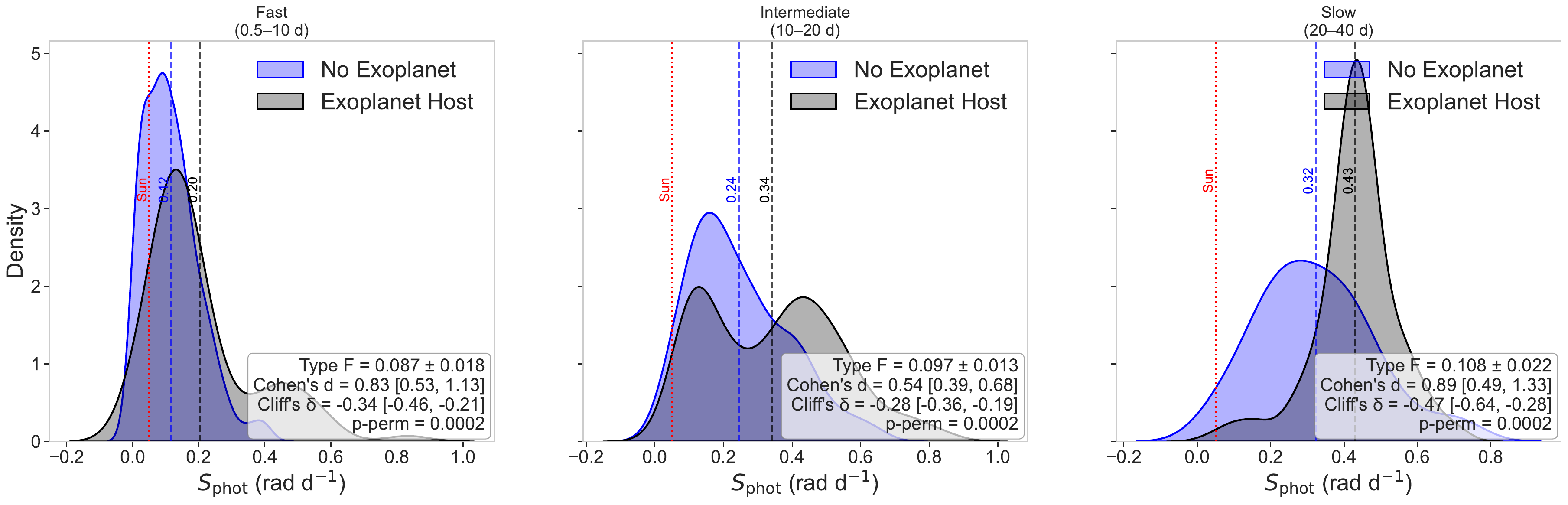}
\caption{
Distributions of $S_{\rm phot}$ and rotation period for stars with (gray) and without (blue) detected exoplanets. The upper panels show results of the \textbf{$S_{\rm phot}$} distributions for spectral types M, K, G, and F, while the middle row panels show the rotation distribution for the same spectral types. The lower panels of the \textbf{$S_{\rm phot}$} distributions group stars into fast (0.5--10 d), intermediate (10--20 d), and slow (20--40 d) rotators. Effect sizes are quantified using Cohen's $d$, and non-parametric differences are indicated by Cliff's $\delta$, with associated confidence intervals and permutation test $p$-values shown in each panel. The solar value (red vertical line) is indicated for reference.
}
\label{fig:teff_rot}
\end{figure*}

Stratifying by spectral type (Figure \ref{fig:teff_rot}), planet-hosting stars show systematically higher \textbf{$S_{\rm phot}$} across all classes of effective temperatures, with the largest effect in cooler stars (Cohen's $d = 1.89 \pm 0.13$ for M dwarfs). Heterogeneity is significant ($Q=30.64$, $p=10^{-6}$), indicating that the enhancement depends on spectral type. This spectral-type dependence is consistent with the effective temperature trends visible in Figure~\ref{fig:densidade}. 

Stratifying by rotation period (bottom row of Figure~\ref{fig:teff_rot}), the enhancement is present in all regimes (fast, intermediate, and slow), with a weighted mean effect size of $d=0.60\pm0.10$. Weak evidence for heterogeneity ($p=0.019$) suggests the effect does not strongly depend on rotation rate.

\begin{figure*}
    \centering
    \includegraphics[width=0.45\linewidth]{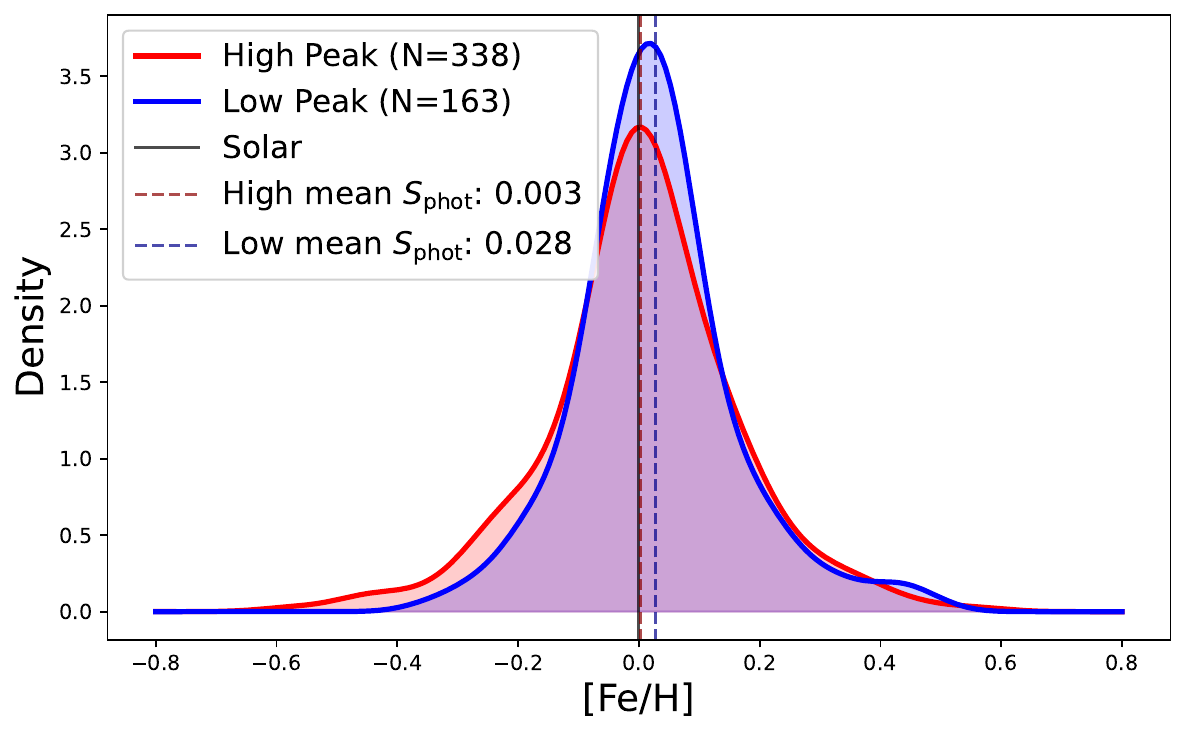}
    \includegraphics[width=0.45\linewidth]{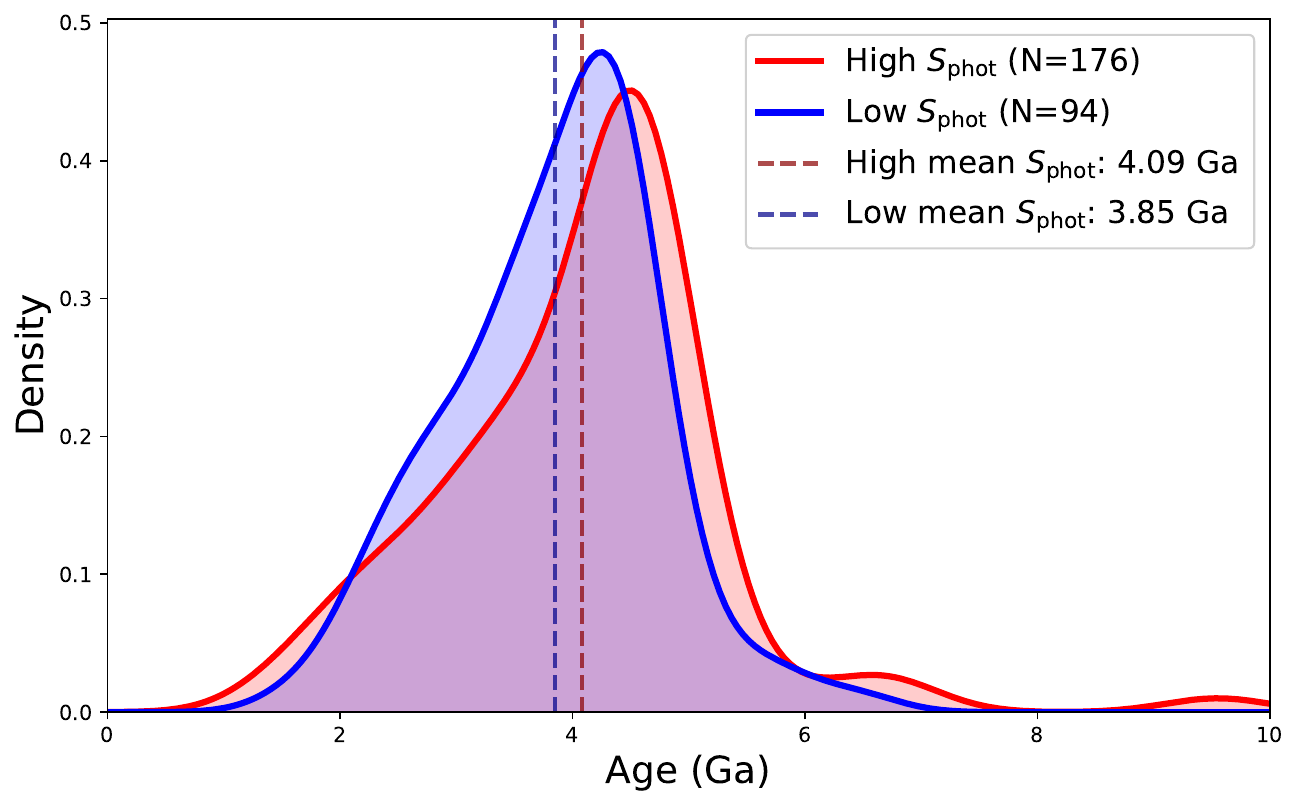}
    \caption{\textbf{Left:} Kernel density estimates revealing the excess of metal-poor stars ([Fe/H]$<-0.2$) in the high-\textbf{$S_{\rm phot}$} peak. 
    Dashed vertical lines indicate the mean metallicity of each peak. The vertical solid line marks solar metallicity ([Fe/H]$=0$). \textbf{Right:} Stellar age distributions, showing substantial overlap between peaks (mean age: $4.09 \pm 1.19$ Gyr for the high peak and $3.85 \pm 0.88$ Gyr for the low peak; difference not statistically significant, $p = 0.088$).}
    \label{fig:metal_age}
\end{figure*}

Metallicity and stellar age distributions (Figure~\ref{fig:metal_age}; Table~\ref{tab:stellar_planetary_deltaomega}) were also examined. The metallicity distribution shows a modest excess of metal-poor stars ([Fe/H] $<-0.2$) in the high-\textbf{$S_{\rm phot}$} peak ($9.8\%$ vs $2.5\%$, $p=0.003$), although the mean [Fe/H] values of the two regimes are statistically indistinguishable. Stellar age likewise does not account for the bimodality: the mean ages of the two peaks are similar ($4.09$ vs $3.85$ Gyr, $p=0.088$), and age exhibits only a very weak correlation with \textbf{$S_{\rm phot}$} within each peak.

\newpage

\bibliography{sample701}{}
\bibliographystyle{aasjournalv7}

\end{document}